\documentclass[12pt,a4paper]{article}

\usepackage{ifthen} 
\newboolean{pdflatex}
\setboolean{pdflatex}{true} 

\newboolean{articletitles}
\setboolean{articletitles}{true} 

\newboolean{uprightparticles}
\setboolean{uprightparticles}{false} 

\newboolean{inbibliography}
\setboolean{inbibliography}{false} 

\usepackage[a4paper,left=27mm,right=27mm,top=35mm,textheight=220mm]{geometry}
\usepackage{microtype}
\usepackage{lineno}  
\usepackage{xspace} 
\usepackage{caption} 

\usepackage{graphicx}  
\usepackage{color}
\usepackage{colortbl}
\graphicspath{{./figs/}} 

\usepackage{amsmath} 
\usepackage{amssymb}
\usepackage{amsfonts}
\usepackage{upgreek} 

\newcommand*\patchAmsMathEnvironmentForLineno[1]{%
    \expandafter\let\csname old#1\expandafter\endcsname\csname #1\endcsname
    \expandafter\let\csname oldend#1\expandafter\endcsname\csname
    end#1\endcsname
    \renewenvironment{#1}%
    {\linenomath\csname old#1\endcsname}%
    {\csname oldend#1\endcsname\endlinenomath}%
}
\newcommand*\patchBothAmsMathEnvironmentsForLineno[1]{%
    \patchAmsMathEnvironmentForLineno{#1}%
    \patchAmsMathEnvironmentForLineno{#1*}%
}
\AtBeginDocument{%
    \patchBothAmsMathEnvironmentsForLineno{equation}%
    \patchBothAmsMathEnvironmentsForLineno{align}%
    \patchBothAmsMathEnvironmentsForLineno{flalign}%
    \patchBothAmsMathEnvironmentsForLineno{alignat}%
    \patchBothAmsMathEnvironmentsForLineno{gather}%
    \patchBothAmsMathEnvironmentsForLineno{multline}%
    \patchBothAmsMathEnvironmentsForLineno{eqnarray}%
}

\usepackage[hidelinks]{hyperref} 
\usepackage[all]{hypcap} 

\def\MagUp {\mbox{\em Mag\kern -0.05em Up}\xspace}

\ifthenelse{\boolean{uprightparticles}}%
{

 \def\Ppi         {\ensuremath{\uppi}\xspace}

 \def\PDelta      {\ensuremath{\Delta}\xspace}                 
 \def\PXi      {\ensuremath{\Xi}\xspace}                 
 \def\PLambda      {\ensuremath{\Lambda}\xspace}                 
 \def\PSigma      {\ensuremath{\Sigma}\xspace}                 
 \def\POmega      {\ensuremath{\Omega}\xspace}                 
 \def\PUpsilon      {\ensuremath{\Upsilon}\xspace}                 
 
 \def\PB      {\ensuremath{\mathrm{B}}\xspace}                 
                  
 \def\PD      {\ensuremath{\mathrm{D}}\xspace}

 \def\PK      {\ensuremath{\mathrm{K}}\xspace}

 \def\Pi      {\ensuremath{\mathrm{i}}\xspace}

}
{

 \def\Ppi         {\ensuremath{\pi}\xspace}

 \mathchardef\PDelta="7101
 \mathchardef\PXi="7104
 \mathchardef\PLambda="7103
 \mathchardef\PSigma="7106
 \mathchardef\POmega="710A
 \mathchardef\PUpsilon="7107
                  
 \def\PB      {\ensuremath{B}\xspace}                 
                  
 \def\PD      {\ensuremath{D}\xspace}

 \def\PK      {\ensuremath{K}\xspace}

 \def\Pi      {\ensuremath{i}\xspace}

}

\makeatletter
\ifcase \@ptsize \relax
  \newcommand{\miniscule}{\@setfontsize\miniscule{4}{5}}
\or
  \newcommand{\miniscule}{\@setfontsize\miniscule{5}{6}}
\or
  \newcommand{\miniscule}{\@setfontsize\miniscule{5}{6}}
\fi
\makeatother

\DeclareRobustCommand{\optbar}[1]{\shortstack{{\miniscule (\rule[.5ex]{1.25em}{.18mm})}
  \\ [-.7ex] $#1$}}

\def\pion   {{\ensuremath{\Ppi}}\xspace}
\def\piz    {{\ensuremath{\pion^0}}\xspace}

\def\pip    {{\ensuremath{\pion^+}}\xspace}
\def\pim    {{\ensuremath{\pion^-}}\xspace}

\def\kaon    {{\ensuremath{\PK}}\xspace}
  \def\Kbar    {{\kern 0.2em\overline{\kern -0.2em \PK}{}}\xspace}

\def\KorKbar    {\kern 0.18em\optbar{\kern -0.18em K}{}\xspace}

\def\Kp      {{\ensuremath{\kaon^+}}\xspace}
\def\Km      {{\ensuremath{\kaon^-}}\xspace}

\def\KS      {{\ensuremath{\kaon^0_{\rm\scriptscriptstyle S}}}\xspace}

  \def\Dbar    {{\kern 0.2em\overline{\kern -0.2em \PD}{}}\xspace}
\def\D       {{\ensuremath{\PD}}\xspace}

\def\DorDbar    {\kern 0.18em\optbar{\kern -0.18em D}{}\xspace}
\def\DtwoorDtwobar {\kern -0.25em\optbar{\kern 0.25em D_2^*}{}\xspace}
\def\Dz      {{\ensuremath{\D^0}}\xspace}
\def\Dzb     {{\ensuremath{\Dbar{}^0}}\xspace}

\def\B       {{\ensuremath{\PB}}\xspace}
\def\Bbar    {{\ensuremath{\kern 0.18em\overline{\kern -0.18em \PB}{}}}\xspace}

\def\BorBbar    {\kern 0.18em\optbar{\kern -0.18em B}{}\xspace}
\def\Bz      {{\ensuremath{\B^0}}\xspace}
\def\Bzb     {{\ensuremath{\Bbar{}^0}}\xspace}
\def\BzorBzbar  {\kern 0.18em\optbar{\kern -0.18em B}{}^0\xspace}
\def\Bu      {{\ensuremath{\B^+}}\xspace}

\def\Bp      {{\ensuremath{\Bu}}\xspace}

  \def\Y#1S{\ensuremath{\PUpsilon{(#1S)}}\xspace}

\def\Lbar        {{\ensuremath{\kern 0.1em\overline{\kern -0.1em\PLambda}}}\xspace}
\def\LorLbar    {\kern 0.18em\optbar{\kern -0.18em \PLambda}{}\xspace}

\def\to                 {\ensuremath{\rightarrow}\xspace}

\def\AT#1     {\ensuremath{A_{\mathrm{T}}^{#1}}\xspace}           

\def\C#1      {\ensuremath{\mathcal{C}_{#1}}\xspace}                       
\def\Cp#1     {\ensuremath{\mathcal{C}_{#1}^{'}}\xspace}                    
\def\Ceff#1   {\ensuremath{\mathcal{C}_{#1}^{\mathrm{(eff)}}}\xspace}        
\def\Cpeff#1  {\ensuremath{\mathcal{C}_{#1}^{'\mathrm{(eff)}}}\xspace}       
\def\Ope#1    {\ensuremath{\mathcal{O}_{#1}}\xspace}                       
\def\Opep#1   {\ensuremath{\mathcal{O}_{#1}^{'}}\xspace}                    

\newcommand{\tev}{\ifthenelse{\boolean{inbibliography}}{\ensuremath{~T\kern -0.05em eV}\xspace}{\ensuremath{\mathrm{\,Te\kern -0.1em V}}}\xspace}
\newcommand{\gev}{\ensuremath{\mathrm{\,Ge\kern -0.1em V}}\xspace}
\newcommand{\mev}{\ensuremath{\mathrm{\,Me\kern -0.1em V}}\xspace}
\newcommand{\kev}{\ensuremath{\mathrm{\,ke\kern -0.1em V}}\xspace}
\newcommand{\ev}{\ensuremath{\mathrm{\,e\kern -0.1em V}}\xspace}
\newcommand{\gevc}{\ensuremath{{\mathrm{\,Ge\kern -0.1em V\!/}c}}\xspace}
\newcommand{\mevc}{\ensuremath{{\mathrm{\,Me\kern -0.1em V\!/}c}}\xspace}
\newcommand{\gevcc}{\ensuremath{{\mathrm{\,Ge\kern -0.1em V\!/}c^2}}\xspace}
\newcommand{\gevgevcccc}{\ensuremath{{\mathrm{\,Ge\kern -0.1em V^2\!/}c^4}}\xspace}
\newcommand{\mevcc}{\ensuremath{{\mathrm{\,Me\kern -0.1em V\!/}c^2}}\xspace}

\def\gsim{{~\raise.15em\hbox{$>$}\kern-.85em
          \lower.35em\hbox{$\sim$}~}\xspace}
\def\lsim{{~\raise.15em\hbox{$<$}\kern-.85em
          \lower.35em\hbox{$\sim$}~}\xspace}

\def\tell1  {TELL1\xspace}
\def\ukl1   {UKL1\xspace}

\usepackage{cite} 
\usepackage{mciteplus}

\usepackage{rotating} 
\usepackage{arydshln} 
\usepackage{relsize} 
\usepackage[hang,flushmargin]{footmisc} 
\usepackage{subcaption}

\begin{document}

\renewcommand{\thefootnote}{\fnsymbol{footnote}}
\setcounter{footnote}{1}

\begin{titlepage}
    \pagenumbering{roman}

    {\bf\boldmath\huge
        \begin{center}
            Dalitz plot binning optimization for the model-independent measurement of the CKM angle $\beta$ with $\Bz \to D\pip\pim$, $D \to \KS\pip\pim$ decays
        \end{center}
    }

    \vspace*{2.0cm}

    \begin{center}
        A.~Li$^1$, W.~Qian$^2$, B.~Shi$^2$, A.~Xu$^3$, L.~Zhang$^1$
        \bigskip\\
        {\it\footnotesize
        $ ^1$ Tsinghua University, Beijing, China\\
        $ ^2$ University of Chinese Academy of Sciences, Beijing, China\\
        $ ^3$ INFN Sezione di Pisa, Pisa, Italy\\
        }
    \end{center}

    \vspace{\fill}

    \begin{abstract}
        \noindent
        A model-independent measurement of the CKM angle $\beta$ can be performed using a time-dependent analysis of $\Bz \to D\pip\pim$, $D \to \KS\pip\pim$ decays, where the Dalitz plots of both the $B$ and $D$ decays are divided into bins.
        The choice of binning scheme for the $\Bz \to D\pip\pim$ Dalitz plot directly impacts the statistical sensitivity to $\beta$.
        This paper presents an optimization of this binning using a Fisher information-based quality factor $Q^2$ and simulated annealing.
        The optimized binning improves the sensitivity to $\beta$ by approximately $10\%$ relative to the equal-phase-difference baseline, as validated by toy Monte Carlo studies without experimental effects.
        The robustness of this improvement against decay time acceptance, Dalitz plot efficiency, Dalitz plot resolution, and background contributions is also assessed.
    \end{abstract}

    \vspace{\fill}

\end{titlepage}

\renewcommand{\thefootnote}{\arabic{footnote}}
\setcounter{footnote}{0}

\pagestyle{plain} 
\setcounter{page}{1}
\pagenumbering{arabic}

\section{Introduction}
\label{sec:intro}

Within the Standard Model, $CP$ violation arises from the complex phase of the Cabibbo-Kobayashi-Maskawa (CKM) quark mixing matrix~\cite{Cabibbo:1963yz,Kobayashi:1973fv}.
The angle $\beta \equiv \arg\left[-V_{cd}V_{cb}^*/(V_{td}V_{tb}^*)\right]$ of the Unitarity Triangle is a fundamental parameter that quantifies $CP$ violation in the $B^0$ system.
The most precise measurements of $\sin2\beta$ come from $b \to c\bar{c}s$ transitions such as $\Bz \to J/\psi \KS$ decays, giving a world average of $\sin2\beta = 0.710 \pm 0.011$~\cite{HFLAV}.
Although penguin contributions with a different weak phase are suppressed in these decays, they introduce a theoretical uncertainty in the extraction of $\beta$ that becomes increasingly important as the experimental precision improves~\cite{Frings:2015eva}.
Independent measurements using decays governed by the tree-level $b \to c\bar{u}d$ transition are therefore particularly valuable: these decays receive no penguin contributions and provide a theoretically clean determination of $\beta$.
Such measurements provide an independent constraint on the Unitarity Triangle and a complementary test of the Standard Model through comparison with $\beta$ determined from $b \to c\bar{c}s$ transitions~\cite{Grossman:1996:decay}.

A natural starting point for a tree-level measurement is $\Bz \to D_{CP}\piz$, where $D_{CP}$ denotes a neutral $D$ meson reconstructed in a $CP$ eigenstate: its time-dependent $CP$ asymmetry gives access to $\sin2\beta$~\cite{Bondar:2005:beta}.
Using $D \to \KS\pip\pim$ instead provides sensitivity to both $\sin2\beta$ and $\cos2\beta$ through the variation of the strong-phase difference across the $D$ Dalitz plot, as proposed in Ref.~\cite{Bondar:2005:beta}.
Access to $\cos2\beta$ provides the additional benefit of resolving the two-fold ambiguity between $\beta$ and $\frac{\pi}{2} - \beta$ left by measurements of $\sin2\beta$ alone.
A joint analysis by the BaBar and Belle collaborations~\cite{Belle:2018:ambiguity,Belle:2018} of $\Bz \to D^{(*)} h^0$ ($h^0 = \piz, \eta, \omega$) decays with $D \to \KS\pip\pim$ has already resolved this ambiguity with more than $5\sigma$ significance. This measurement relies on a $D$ decay amplitude model, introducing model-dependent systematic uncertainties.
Belle also performed a model-independent measurement using a binned $D$ Dalitz plot analysis with externally measured strong-phase parameters~\cite{Belle:2016:beta}.

At LHCb, reconstructing neutral particles such as $\piz$ and $\eta$ makes the $\Bz \to D^{(*)}h^0$ channels experimentally challenging.
This motivates the use of $\Bz \to D\pip\pim$ for a tree-level determination of $\beta$: with $D \to \KS\pip\pim$ and $\KS \to \pip\pim$, or with $D$ decays to $CP$ eigenstates containing only charged particles, the decay chain can be reconstructed entirely from charged tracks~\cite{Latham:2008,Bondar:2018:beta}.
Latham and Gershon~\cite{Latham:2008} proposed a time-dependent Dalitz plot analysis of $\Bz \to D\pip\pim$ decays with the $D$ meson reconstructed in the $CP$-even final states $D \to \Kp\Km$ and $D \to \pip\pim$ to determine both $\sin2\beta$ and $\cos2\beta$.
The method requires information on the variation of the $\Bz \to D\pip\pim$ decay amplitude across its Dalitz plot, which was described using an amplitude model in the original proposal.

Bondar, Kuzmin and Vorobyev~\cite{Bondar:2018:beta} extended the model-independent binned approach to the double-Dalitz decay $\Bz \to D\pip\pim$, $D \to \KS\pip\pim$.
In this method, the Dalitz plots of both the $B$ and $D$ decays are divided into bins, and the hadronic parameters describing the $B$ decay in each bin are determined from the data simultaneously with $\beta$, eliminating the need for a $B$ decay amplitude model.
The same work considered a joint analysis of this double-Dalitz channel and $\Bz \to D_{CP}\pip\pim$, sharing the $B$-decay hadronic parameters between the samples to constrain them together with $\beta$.
Alternatively, the measured $B$-decay strong-phase parameters $c_j$ and $s_j$ could subsequently be used as external inputs to the $D \to \Kp\Km$ and $D \to \pip\pim$ analysis, allowing it to be performed without a $B$-decay amplitude model.
The $D$ decay hadronic parameters are taken from external measurements using quantum-correlated $\Dz\Dzb$ pairs at charm factories~\cite{Libby:2010nu,BESIII:2020:kspipi,BESIII:2025:kspipi}, and are expected to be further improved by BESIII using its full $20\text{ fb}^{-1}$ $\psi(3770)$ data sample~\cite{BESIII:2026:qc}.
This binned approach was developed for $\Bp \to D\Kp$ decays~\cite{Giri:2003ty,Bondar:2005ki} and later extended to $\Bz \to D\Kp\pim$ decays for the measurement of $\gamma$~\cite{Gershon:2009qr,Craik:2018}. In the $\gamma$ measurement, the $D \to \KS\pip\pim$ Dalitz plot binning is optimized by maximizing a quality factor $Q^2$ derived from the $D$ decay strong-phase parameters $(C_i, S_i)$, which quantifies the sensitivity to the $CP$-violating observables~\cite{Bondar:2008hh}. The resulting ``optimal'' scheme with $\mathcal{N}=8$ ($16$ bins) is used as external input in this analysis.
The binned approach using $D \to \KS\pip\pim$ was shown to be feasible with the data samples expected at Belle II and the LHCb upgrade, where a statistical precision of approximately $1^\circ$ on $\beta$ can be reached, and systematic biases from neglected sub-leading amplitudes and charm mixing are at the sub-degree level~\cite{Bondar:2018:beta}.

A key ingredient in maximizing the sensitivity of the binned analysis is the choice of binning scheme for the $\Bz \to D\pip\pim$ Dalitz plot.
Previous studies of model-independent $\gamma$ measurements have shown that optimizing a binning quality factor can improve the sensitivity relative to equal-phase-difference binning~\cite{Bondar:2008hh}.
In the present study, a binning quality factor is constructed from the Fisher information so that it directly quantifies the statistical sensitivity.
This paper presents a dedicated study of the binning optimization for the $\Bz \to D\pip\pim$, $D \to \KS\pip\pim$ analysis.
The sensitivity improvement from the optimization is quantified and validated through toy Monte Carlo studies, and the effects of experimental conditions such as decay time acceptance, Dalitz plot efficiency, and the number of bins are examined.

The formalism of the time-dependent binned analysis is reviewed in Sec.~\ref{sec:formalism}.
The binning optimization strategy is presented in Sec.~\ref{sec:strategy}, and the results are discussed in Sec.~\ref{sec:results}.
A summary is given in Sec.~\ref{sec:summary}.

\section{Formalism}
\label{sec:formalism}

Up to the common decay factor $e^{-t/(2\tau_B)}$, the time-dependent amplitude for the decay $\Bz \to D\pip\pim$, with $D \to \KS\pip\pim$, can be written as~\cite{Bondar:2018:beta}
\begin{align}
    \mathcal{A}(t, \mu_+^2, \mu_-^2, m_+^2, m_-^2)
    ={} & \mathcal{A}_{B \to \bar{D}}(\mu_+^2, \mu_-^2) \, \bar{\mathcal{A}}_D(m_+^2, m_-^2) \,
    \cos\!\left(\frac{\Delta m_B t}{2}\right) \nonumber                                            \\
        & {}+ i \, \mathcal{A}_{\bar{B} \to D}(\mu_+^2, \mu_-^2) \, \mathcal{A}_D(m_+^2, m_-^2) \,
    e^{-2i\beta} \sin\!\left(\frac{\Delta m_B t}{2}\right) \,
    \label{eq:amplitude}
\end{align}
where $t$ is the decay time, and $(\mu_+^2, \mu_-^2)$ and $(m_+^2, m_-^2)$ denote the Dalitz plot coordinates of the $\Bz \to D\pip\pim$ and $D \to \KS\pip\pim$ decays, respectively.
Here $\mathcal{A}_{B \to \bar{D}}$ ($\mathcal{A}_{\bar{B} \to D}$) is the amplitude for $\Bz \to \Dzb \pip\pim$ ($\Bzb \to \Dz \pip\pim$), and $\bar{\mathcal{A}}_D$ ($\mathcal{A}_D$) is the amplitude for $\Dzb \to \KS\pip\pim$ ($\Dz \to \KS\pip\pim$).
Neglecting direct $CP$ violation in the $D$ decay, $\mathcal{A}_D(m_+^2, m_-^2) = \bar{\mathcal{A}}_D(m_-^2, m_+^2)$. Charm mixing is also neglected; its effect on the $\beta$ measurement has been shown to be at the sub-degree level~\cite{Bondar:2018:beta}.
The time-dependent probability density function (PDF) is obtained from the squared modulus of the amplitude, accounting for the initial flavour of the $\Bz$ meson and the $\Bz$--$\Bzb$ oscillation.

Following the binned approach, the $\Bz \to D\pip\pim$ and $D \to \KS\pip\pim$ Dalitz plots are both divided into bins symmetrically under the exchange of the respective Dalitz plot coordinates: $2\mathcal{M}$ bins for the $B$ decay (indexed by $j = -\mathcal{M}, \dots, \mathcal{M}$, excluding 0), and $2\mathcal{N}$ bins for the $D$ decay (indexed by $i = -\mathcal{N}, \dots, \mathcal{N}$, excluding 0).
For the $D$ decay, the established ``optimal'' binning scheme with $\mathcal{N}=8$~\cite{Libby:2010nu} is used.

The $B$ decay hadronic parameters are defined for each bin $j = \pm 1, \dots, \pm\mathcal{M}$ as
\begin{eqnarray}
    k_j &\equiv& \int_{\mathcal{B}_j} |\mathcal{A}_{B \to \bar{D}}(\mu_+^2, \mu_-^2)|^2 \, d\mu_+^2 d\mu_-^2 \,, \label{eq:kj} \\
    c_j + i s_j &=& \frac{1}{\sqrt{k_j k_{-j}}}
    \int_{\mathcal{B}_j} \mathcal{A}_{B \to \bar{D}}^*(\mu_+^2, \mu_-^2) \,
    \mathcal{A}_{B \to \bar{D}}(\mu_-^2, \mu_+^2) \, d\mu_+^2 d\mu_-^2 \,,
    \label{eq:cjsj}
\end{eqnarray}
where $\mathcal{B}_j$ denotes the phase-space region of bin $j$, and bin $-j$ is related to bin $j$ by the exchange $\mu_+^2 \leftrightarrow \mu_-^2$. By $CP$ symmetry, $\mathcal{A}_{\bar{B} \to D}(\mu_+^2,\mu_-^2)=\mathcal{A}_{B \to \bar{D}}(\mu_-^2,\mu_+^2)$ and $\int_{\mathcal{B}_j} |\mathcal{A}_{\bar{B} \to D}|^2 = k_{-j}$. The parameters satisfy the physical constraint $c_j^2 + s_j^2 \leq 1$.

Similarly, the $D$ decay hadronic parameters are
\begin{eqnarray}
    K_i &=& \int_{\mathcal{D}_i} |\mathcal{A}_D(m_+^2, m_-^2)|^2 \, dm_+^2 dm_-^2 \,, \label{eq:Ki} \\
    C_i + i S_i &=& \frac{1}{\sqrt{K_i K_{-i}}}
    \int_{\mathcal{D}_i} \mathcal{A}_D^*(m_+^2, m_-^2) \,
    \mathcal{A}_D(m_-^2, m_+^2) \, dm_+^2 dm_-^2 \,,
    \label{eq:CiSi}
\end{eqnarray}
with $C_i^2 + S_i^2 \leq 1$.
For the symmetric binning, the strong-phase parameters obey $c_{-j}=c_j$, $s_{-j}=-s_j$, $C_{-i}=C_i$, and $S_{-i}=-S_i$.

After binning, the time-dependent PDF for a candidate with reconstructed decay time $t$, initial-flavour tag $q$ ($q = +1$ when tagged as a $\Bz$ and $-1$ when tagged as a $\Bzb$), and falling into $B$ bin $j$ and $D$ bin $i$, is
\begin{equation}
    p_{ij}(t, q) \propto e^{-t/\tau_B} \Big[
        U_{ij} + q(1 - 2\omega)\big(
        D_{ij} \cos(\Delta m_B t) + F_{ij} \sin(\Delta m_B t)
        \big)
        \Big] \,,
    \label{eq:pdf}
\end{equation}
where $\tau_B$ is the $\Bz$ lifetime, $\Delta m_B$ is the $\Bz$--$\Bzb$ oscillation frequency, and $\omega$ is the flavour mistag probability.
The numerical study uses the positive decay-time convention $t\geq0$, appropriate to a hadron-collider measurement.
For an asymmetric-energy $B$ factory, the corresponding expression uses the signed decay-time difference $\Delta t$ and the factor $e^{-|\Delta t|/\tau_B}$.
The coefficients $U_{ij}$, $D_{ij}$, and $F_{ij}$ are expressed in terms of the binned hadronic parameters as
\begin{eqnarray}
    U_{ij} &=& \tfrac{1}{2} \big( K_{-i} k_j + K_i k_{-j} \big) \,, \\
    D_{ij} &=& \tfrac{1}{2} \big( K_{-i} k_j - K_i k_{-j} \big) \,, \\
    F_{ij} &=& -\sqrt{K_i K_{-i} k_j k_{-j}} \,
    \big[ (C_i s_j - S_i c_j) \cos 2\beta - (C_i c_j + S_i s_j) \sin 2\beta \big] \,.
    \label{eq:Fij}
\end{eqnarray}
Equation~(\ref{eq:pdf}) is the basis for the maximum likelihood fit from which $\beta$ and the $B$ decay hadronic parameters $k_j$, $k_{-j}$, $c_j$, and $s_j$ are determined.
The sensitivity to $\beta$ depends on the magnitude of the interference term $F_{ij}$, which in turn depends on the binning scheme through the parameters $k_j$, $c_j$, and $s_j$.

\section{Optimization strategy}
\label{sec:strategy}

We consider a measurement in which the hadronic parameters are treated as known and only $\beta$ is floated.
In this configuration, the statistical uncertainty on $\beta$ is bounded by the Cram\'er--Rao inequality $\sigma_{\beta}^2 \geq I^{-1}$, where
\begin{equation}
    I = -\mathbb{E}\!\left[ \frac{\partial^2 \ln L}{\partial \beta^2} \right]
    \label{eq:fisher}
\end{equation}
is the Fisher information for $\beta$, $\mathbb{E}$ denotes the expectation value over data distributed according to the PDF of Eq.~(\ref{eq:pdf}), and $L$ is the corresponding likelihood~\cite{Rao:1945}.
The same fit configuration is used in the toy verification in Sec.~\ref{sec:toy_verification}.
If $\beta$ and the hadronic parameters are instead determined simultaneously, the corresponding bound is given by the $(\beta,\beta)$ element of the inverse Fisher information matrix.
In the large-sample limit, the maximum likelihood estimator saturates the bound, so $\sigma_{\rm stat} \simeq 1 / \sqrt{I}$.
To quantify the impact of binning on the sensitivity, we define a binning quality factor
\begin{equation}
    Q^2 = \frac{I_{\rm binned}}{I_{\rm unbinned}} \,,
    \label{eq:Q2_def}
\end{equation}
which represents the fraction of the Fisher information retained after binning the Dalitz plot.
A value of $Q^2$ close to unity indicates that the binning introduces minimal loss of sensitivity.
The goal of the optimization is to find the binning scheme that maximizes $Q^2$.

\subsection{Time-integrated method}

As a simpler reference, we first consider a time-integrated (TI) parameterization of the amplitude, analogous to the approach used for $\gamma$ measurements~\cite{Gershon:2009qr}.
The amplitude is written as
\begin{equation}
    A = A_1 + (x + iy) A_2 \,,
\end{equation}
where $A_1 \equiv \mathcal{A}_{B \to \bar{D}} \cdot \bar{\mathcal{A}}_D$ and $A_2 \equiv \mathcal{A}_{\bar{B} \to D} \cdot \mathcal{A}_D$.
The complex parameter $x + iy$ encodes the effective $CP$-violating interference strength between the two amplitude chains, analogous to $x_\pm + iy_\pm = r_B e^{i(\delta_B \pm \gamma)}$ in the $\gamma$ measurement~\cite{Gershon:2009qr}. In the time-integrated approximation, the time-dependent phase factors are treated as constant, and the $CP$-violating structure is captured by $x + iy$ alone.
For the parameter vector $\boldsymbol{\theta} = (x,y)$, the information matrix is
\begin{equation}
    I_{ab} = \int \frac{1}{p(\Phi;\boldsymbol{\theta})}
    \frac{\partial p(\Phi;\boldsymbol{\theta})}{\partial \theta_a}
    \frac{\partial p(\Phi;\boldsymbol{\theta})}{\partial \theta_b}
    \,d\Phi \, ,
    \qquad \theta_a,\theta_b \in \{x,y\} \, ,
    \label{eq:fisher_TI}
\end{equation}
where $p(\Phi;\boldsymbol{\theta}) = |A|^2$ is the event density over the combined phase space $\Phi=(\Phi_B,\Phi_D)$.
The sum of the diagonal elements, $I_{xx}+I_{yy}$, quantifies the combined sensitivity to the two Cartesian parameters, and the TI quality factor is defined as
\begin{equation}
    Q^2_{\rm TI} =
    \frac{(I_{xx}+I_{yy})_{\rm binned}}
    {(I_{xx}+I_{yy})_{\rm unbinned}} \, .
    \label{eq:Q2_TI_info}
\end{equation}
This quantity depends only weakly on the actual values of $(x,y)$ and is evaluated at $(x,y)=(0,0)$~\cite{Bondar:2008hh}. At this point, $p=|A_1|^2$, $\partial p/\partial x=2\operatorname{Re}(A_1^*A_2)$, and $\partial p/\partial y=-2\operatorname{Im}(A_1^*A_2)$, making explicit how the interference between the two amplitude chains determines the sensitivity.
Integrating over the $B$ and $D$ Dalitz plots, the quality factor takes the factorized form
\begin{equation}
    Q^2_{\rm TI} = \frac{\sum_j (c_j^2 + s_j^2) k_j}{\sum_j k_j}
    \cdot \frac{\sum_i (C_i^2 + S_i^2) K_i}{\sum_i K_i} \,,
    \label{eq:Q2_TI}
\end{equation}
where the first factor depends only on the $B$ decay binning and the second only on the $D$ decay binning.
Since the $D$ decay binning is fixed from external measurements, the optimization reduces to maximizing the $B$ decay factor independently.
The TI method has the advantage of a simple, analytic objective function but does not account for the time dependence or flavour tagging, which are essential to the $\beta$ measurement.

\subsection{Time-dependent method}

The time-dependent (TD) method uses the full PDF of Eq.~(\ref{eq:pdf}) to construct the Fisher information for the parameter $\beta$.
The quality factor is
\begin{equation}
    Q^2_{\rm TD} = \frac{\int dt \sum_{ij} \sum_q \frac{1}{p_{ij}} \left( \frac{\partial p_{ij}}{\partial \beta} \right)^2}
    {\int dt \int d\Phi \sum_q \frac{1}{p} \left( \frac{\partial p}{\partial \beta} \right)^2} \,,
    \label{eq:Q2_TD}
\end{equation}
where the numerator sums the Fisher information over all binned cells after integrating over decay time $t$ and summing over initial-flavour tags $q$, while the denominator gives the corresponding unbinned integral over the full phase space $d\Phi = d\Phi_B \, d\Phi_D$.
The integration over decay time is performed numerically over the representative range $0.3 < t < 14\,\mathrm{ps}$.
The binning-independent denominator is omitted in the numerical objective, since it cancels in ratios of $Q$ values.
In contrast to the TI method, the TD quality factor does not factorize into separate $B$ and $D$ contributions, and the optimization must consider the full combined $(B, D)$ binning.
However, since the $D$ binning is held fixed at the established optimal scheme, the optimization still varies only the $B$-decay bin assignments.

The TD method provides the most complete description of the $\beta$ analysis, as it incorporates the time-dependent $\Bz$--$\Bzb$ oscillation structure, the effect of flavour tagging (including the mistag probability $\omega$), and the dependence on the true value of $\beta$.
It is therefore expected to give the most reliable prediction of the sensitivity improvement from binning optimization.

\subsection{Simulated annealing optimization}
\label{sec:annealing}

\begin{figure}[htb]
    \centering
    \includegraphics[width=0.9\linewidth]{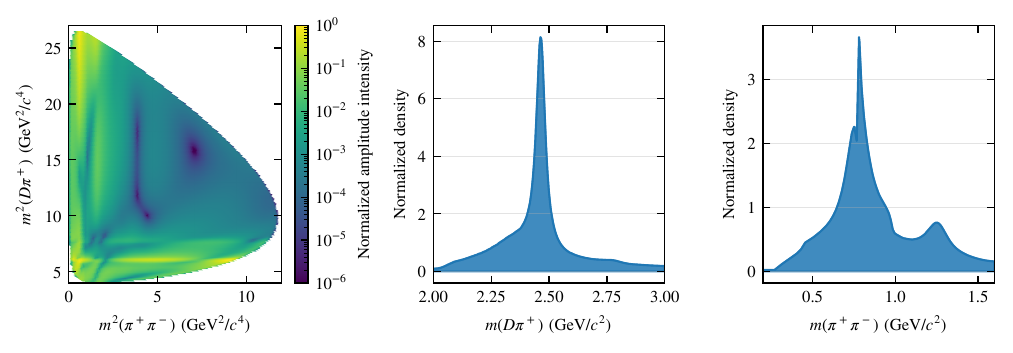}
    \caption{
        Distribution of $\Bz \to \Dzb\pip\pim$ decays from the amplitude model of Ref.~\cite{LHCb:2015}, used to calculate the $B$ decay hadronic parameters $k_j$, $c_j$, and $s_j$.
    }
    \label{fig:hist_dalitz_B}
\end{figure}

To search for a binning scheme with high $Q^2$, a simulated annealing algorithm~\cite{Kirkpatrick:1983} is employed.
The $B$ decay hadronic parameters $k_j$, $c_j$, and $s_j$ are calculated using the publicly available amplitude model of $\Bz \to \Dzb\pip\pim$ from Ref.~\cite{LHCb:2015}. Figure~\ref{fig:hist_dalitz_B} shows the $B$ Dalitz plot distribution predicted by this model.

The optimization starts from an equal-phase-difference binning, where the $\Bz \to D\pip\pim$ Dalitz plot is divided such that each bin spans an equal range of the strong phase difference $\delta_B(\mu_+^2, \mu_-^2) = \arg(\mathcal{A}_{B \to \bar{D}}^* \, \mathcal{A}_{\bar{B} \to D})$.

At each iteration, a new binning is proposed by randomly selecting a symmetric pair of pixels in the discretized Dalitz plot and reassigning their bin labels.
With probability 60\%, the new label is drawn uniformly from all allowed labels; otherwise, the previous label is shifted by $+1$ or $-1$ with equal probability, with a uniform draw used if the shifted label falls outside the allowed range.
The $\pm1$ proposals favor changes between neighboring bin labels and are intended to encourage the formation of contiguous bin regions.
Attempts that leave the labels unchanged are skipped.
The acceptance decision uses the relative change in the objective,
\begin{equation}
    \delta_{\rm rel} = \frac{Q^2_{\rm new} - Q^2_{\rm old}}{Q^2_{\rm old}} \, .
    \label{eq:annealing_relative_change}
\end{equation}
The proposed binning is always accepted if $\delta_{\rm rel}\geq 0$; otherwise, it is accepted with probability $\exp(\delta_{\rm rel}/T)$ for $T>0$ and rejected at $T=0$.
The temperature decreases according to the cooling schedule $T = T_0[1 - n/(N_{\rm total}-1)]^k$.
Here, $T_0$ is the initial temperature, $N_{\rm total}$ is the total number of iterations in an optimization run, $n=0,\ldots,N_{\rm total}-1$ is the iteration index, and $k$ controls the cooling rate.
As the temperature approaches zero, the algorithm settles on a high-$Q^2$ binning scheme for that optimization run.
The binning with the largest $Q^2$ encountered during the optimization is retained.
The cooling parameters are listed in Table~\ref{tab:annealing_parameters}.
These choices aim to provide a sufficiently high initial acceptance rate for broad exploration of binning schemes, followed by a gradual decrease that favors convergence to a stable, high-$Q^2$ solution.

\begin{table}[htb]
    \centering
    \caption{Cooling parameters for the binning optimization.}
    \label{tab:annealing_parameters}
    \begin{tabular}{lccc}
        \hline
        Optimization          & $T_0$            & $N_{\rm total}$  & $k$ \\
        \hline
        TI                    & $2\times10^{-5}$ & $2.4\times10^7$  & 20  \\
        TD without background & $10^{-4}$        & $4.32\times10^6$ & 25  \\
        TD with background    & $5\times10^{-6}$ & $1.68\times10^6$ & 50  \\
        \hline
    \end{tabular}
\end{table}

The TD quality factor is evaluated assuming $\beta = 22^\circ$, close to the current world average, and a mistag probability $\omega = 0.36$. The $\Bz$ lifetime $\tau_B$ and oscillation frequency $\Delta m_B$ are fixed to values consistent with their world averages~\cite{HFLAV}.

The baseline against which improvements are measured is the equal-phase-difference binning, which serves as a physically motivated reference before any optimization is applied.

\begin{figure}[htb]
    \centering
    \includegraphics[width=0.8\textwidth]{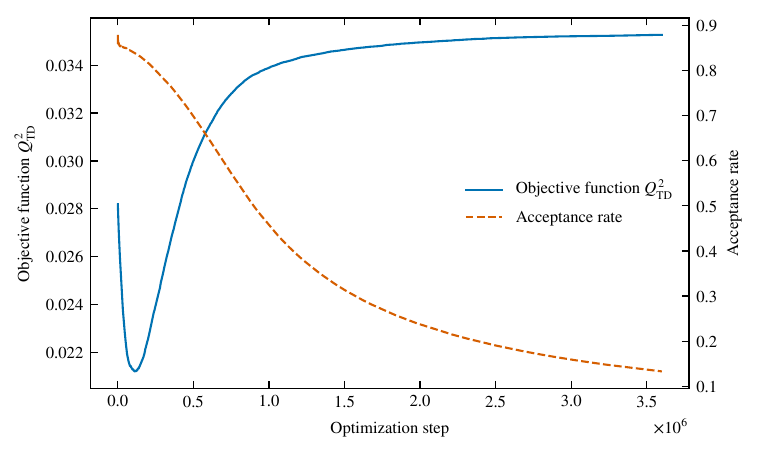}
    \caption{
    Example of convergence of the time-dependent binning optimization for $\mathcal{M}=8$, with $T_0=10^{-4}$, $N_{\rm total}=3.6\times10^6$, and $k=25$.
    The objective function, proportional to $Q^2_{\rm TD}$ (blue), and the cumulative proposal acceptance rate (orange) are shown as functions of the optimization step.
    After initial high-temperature fluctuations, the objective function rises toward a plateau, while the cumulative acceptance rate decreases as the simulated annealing temperature is reduced.
    }
    \label{fig:optimization_stats}
\end{figure}

The convergence behavior of the simulated annealing optimization for the TD method is illustrated in Fig.~\ref{fig:optimization_stats}.
After initial high-temperature fluctuations, the objective function $Q^2_{\rm TD}$ (left axis) rises as the algorithm discovers binning schemes with improved sensitivity, while the cumulative acceptance rate (right axis) gives the fraction of label-changing proposals accepted up to each step.
In the early stages, the high temperature allows many proposals to be accepted, including those that temporarily reduce $Q^2$.
This helps the algorithm explore a broad range of binning configurations and escape local maxima.
As the temperature decreases, moves that lower the objective function become less likely to be accepted.
The eventual plateau in $Q^2$ shows that the objective changes little toward the end of this run.

\section{Optimization results}
\label{sec:results}

The binning optimization is performed using both the TI and TD methods described in Sec.~\ref{sec:strategy}.
The $D$ Dalitz plot is binned using the optimal scheme with $\mathcal{N}=8$ ($16$ bins), and this binning is held fixed throughout the optimization.
The baseline against which improvements are measured is the equal-phase-difference binning of the $B$ Dalitz plot.
\subsection{Optimized binning and sensitivity improvement \boldmath($\mathcal{M}=8$)}

\begin{figure}[htb]
    \centering
    \begin{subfigure}[T]{0.95\textwidth}
        \centering
        \includegraphics[width=1\textwidth]{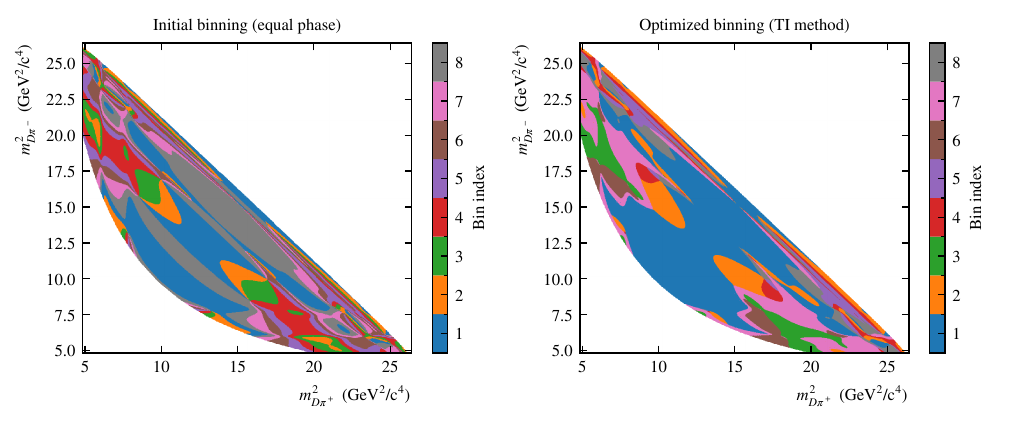}
        \caption{TI method}
    \end{subfigure}
    \begin{subfigure}[T]{0.95\textwidth}
        \centering
        \includegraphics[width=1\textwidth]{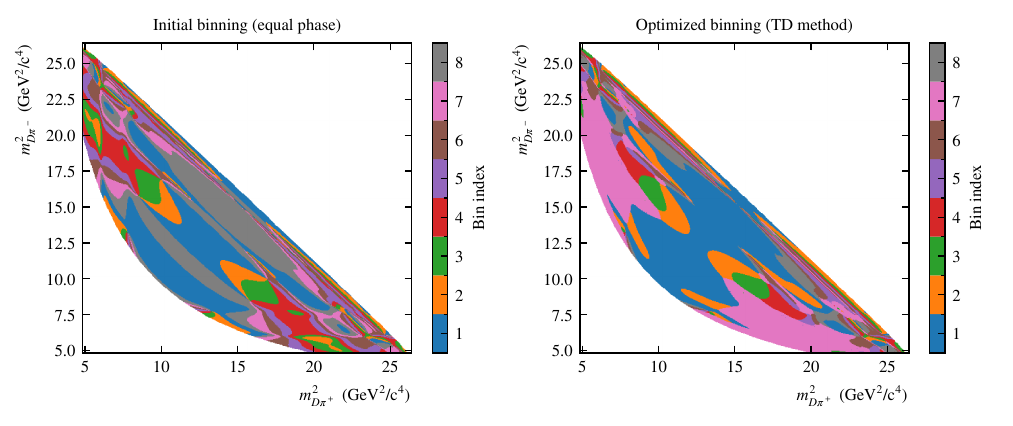}
        \caption{TD method}
    \end{subfigure}
    \caption{
        Comparison of the initial equal-phase-difference binning (left) and the optimized binning (right) for the (a) TI and (b) TD methods with $\mathcal{M}=8$.
        Colours indicate the absolute bin index $|j|$; paired bins $j$ and $-j$ share the same colour.
    }
    \label{fig:binning_evolution}
\end{figure}

The nominal results are obtained with $\mathcal{M}=8$ in the $\Bz \to D\pip\pim$ Dalitz plot (i.e.\ $2\mathcal{M}=16$ bins symmetric under $\mu_+^2 \leftrightarrow \mu_-^2$).
Figure~\ref{fig:binning_evolution} compares the initial equal-phase-difference binning with the optimized binning for the TI and TD methods.

\begin{figure}[htb]
    \centering
    \begin{subfigure}[T]{0.47\textwidth}
        \centering
        \includegraphics[width=1\textwidth]{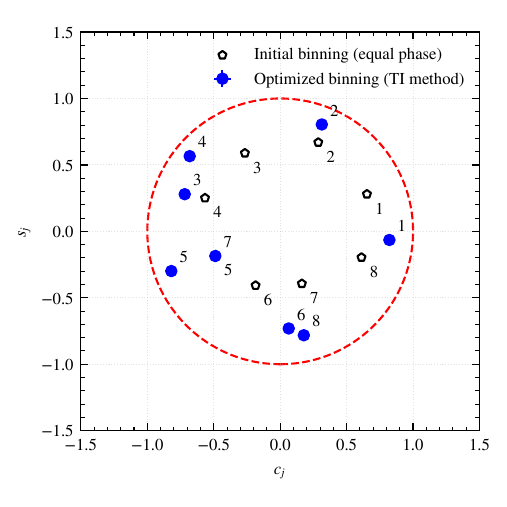}
        \caption{TI method}
    \end{subfigure}
    \begin{subfigure}[T]{0.47\textwidth}
        \centering
        \includegraphics[width=1\textwidth]{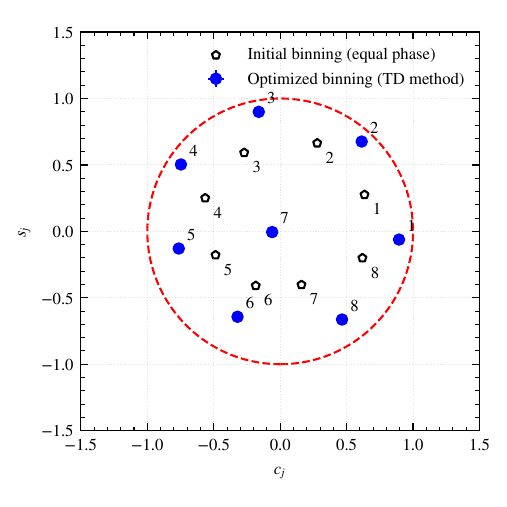}
        \caption{TD method}
    \end{subfigure}
    \caption{
        Distributions of the $B$ decay hadronic parameters $c_j$ and $s_j$ for the initial equal-phase-difference and optimized binning schemes with $\mathcal{M}=8$, using the (a) TI and (b) TD methods.
        Each point represents one positive-index bin; the parameters of the paired negative-index bin obey $c_{-j}=c_j$ and $s_{-j}=-s_j$.
        The dashed circle indicates the physical boundary $c_j^2 + s_j^2 = 1$.
    }
    \label{fig:cs}
\end{figure}

Figure~\ref{fig:cs} compares the $B$ decay hadronic parameters $c_j$ and $s_j$ in the complex plane before and after optimization.
Each point corresponds to one positive-index $B$ bin, and the parameters of the paired negative-index bin obey $c_{-j}=c_j$ and $s_{-j}=-s_j$.
The distance of each point from the origin is $\sqrt{c_j^2 + s_j^2}$.
For the TI result, the optimization tends to move the points toward the unit circle ($c_j^2 + s_j^2 \approx 1$), thereby directly increasing $Q^2_{\rm TI}$ according to Eq.~(\ref{eq:Q2_TI}). The TD result places most points near the unit circle, while one bin remains close to the origin. This occurs because some regions of the Dalitz plot have little sensitivity to $\beta$. Grouping much of this low-sensitivity phase space into one bin avoids diluting the sensitivity of the other bins.

\begin{figure}[htb]
    \centering
    \begin{subfigure}[T]{0.47\textwidth}
        \centering
        \includegraphics[width=1\textwidth]{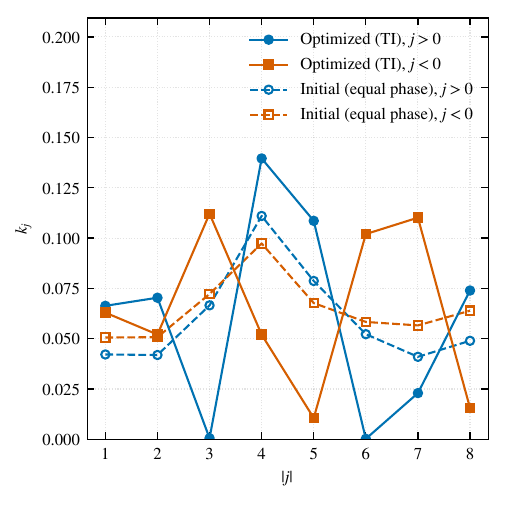}
        \caption{TI method}
    \end{subfigure}
    \begin{subfigure}[T]{0.47\textwidth}
        \centering
        \includegraphics[width=1\textwidth]{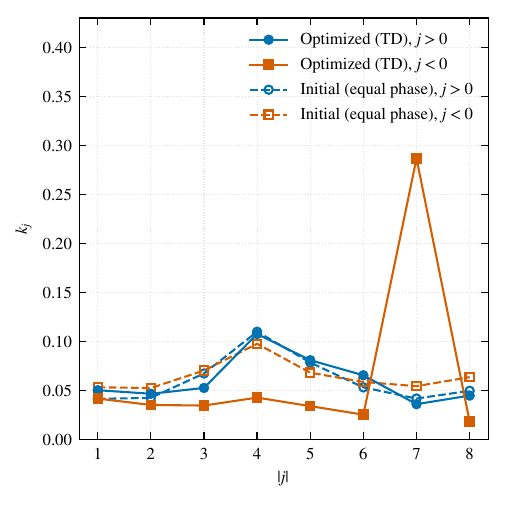}
        \caption{TD method}
    \end{subfigure}
    \caption{
        Distributions of the $B$ decay hadronic parameters $k_j$ for the initial equal-phase-difference and optimized binning schemes with $\mathcal{M}=8$, using the (a) TI and (b) TD methods.
        Blue circles and orange squares indicate positive- and negative-index bins, respectively.
        Open markers with dashed lines denote the initial binning, and filled markers with solid lines denote the optimized binning.
    }
    \label{fig:k}
\end{figure}

Figure~\ref{fig:k} compares the $B$ decay hadronic parameters $k_j$ for the initial equal-phase-difference and optimized binnings.
The values are normalized such that $\sum_{j=1}^{\mathcal{M}} (k_j + k_{-j}) = 1$.

The uncertainty improvement relative to the equal-phase-difference baseline is quantified in Table~\ref{tab:results}.
The improvement is defined as $(\sigma_{\rm baseline} - \sigma_{\rm opt}) / \sigma_{\rm baseline} = 1 - Q_{\rm baseline} / Q_{\rm opt}$.
For each method, two values are reported: the improvement evaluated using the TI quality factor ($Q_{\rm TI}$), and the improvement evaluated using the TD quality factor ($Q_{\rm TD}$).
The TD quality factor is the relevant metric for this study because it is derived from the full time-dependent PDF used in the $\beta$ fit. Its prediction is validated by the toy Monte Carlo studies presented in Sec.~\ref{sec:toy_verification}.

\begin{table}[htb]
    \centering
    \caption{
        Uncertainty improvement (\%) of the optimized binning relative to the equal-phase-difference baseline for $\mathcal{M}=8$.
        $Q_{\rm TI}$ is evaluated using the TI quality factor; $Q_{\rm TD}$ is evaluated using the TD quality factor.
        Both evaluations use the same binned parameters calculated without experimental effects.
    }
    \label{tab:results}
    \begin{tabular}{lcc}
        \hline
        Method                      & $Q_{\rm TI}$ (\%) & $Q_{\rm TD}$ (\%) \\
        \hline
        TI-optimized binning scheme & 23.9              & 5.0               \\
        TD-optimized binning scheme & 13.4              & 9.7               \\
        \hline
    \end{tabular}
\end{table}

The TI method predicts a substantial uncertainty improvement using its own quality factor ($\sim 24\%$), but this translates to a more modest gain when evaluated with the more realistic TD metric ($\sim 5\%$).
The TD method, which directly optimizes the relevant Fisher information, yields a sensitivity improvement of approximately $10\%$ over the baseline.

For the nominal TD optimization at $\mathcal{M}=8$, the relative spread between optimized expected uncertainties, $(\sigma_{\beta,\max}-\sigma_{\beta,\min})/\sigma_{\beta,\min}$ with $\sigma_\beta\propto1/Q_{\rm TD}$, is below $0.01\%$ across ten random seeds with equal-phase-difference initialization and below $1\%$ when random initial binnings are also included.
We adopt a representative of the family containing the largest $Q_{\rm TD}$ found in these scans; the family members have very similar binning schemes, hadronic parameters and $Q_{\rm TD}$ values.

\subsection{Effects on the optimization}
\label{sec:effects}

\subsubsection{Dependence on the number of bins $\mathcal{M}$}

To study the dependence on $\mathcal{M}$, the binning optimization is repeated for $\mathcal{M} = 2, 4, \dots, 20$, and the resulting binning schemes are evaluated using toy pseudoexperiments.
For this study, an idealized toy configuration is used in which decay time acceptance and Dalitz plot efficiency are not applied, in order to isolate the dependence on $\mathcal{M}$ alone. A benchmark sample size of approximately $13\,000$ signal events is assumed in each pseudoexperiment.
For each value of $\mathcal{M}$, the uncertainty reported by the fit is recorded for every pseudoexperiment, and the $\beta$ uncertainty is taken as its mean over the toy ensemble, for both the equal-phase-difference baseline and the TD-optimized binning.

\begin{figure}[htbp]
    \centering
    \includegraphics[width=0.9\linewidth]{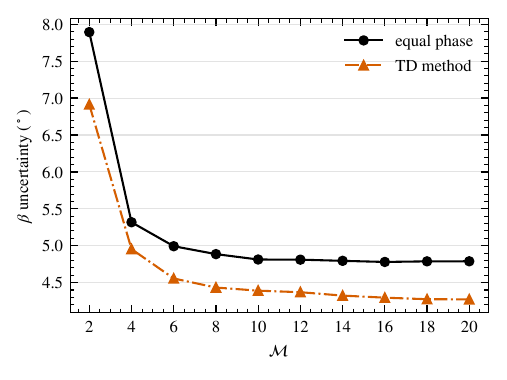}
    \caption{
        Mean fit-reported $\beta$ uncertainty as a function of $\mathcal{M}$ for the equal-phase-difference baseline and the TD-optimized binning, obtained from toy pseudoexperiments with approximately $13\,000$ signal events and without decay-time-acceptance or Dalitz-efficiency effects.
    }
    \label{fig:M_dependence}
\end{figure}

Figure~\ref{fig:M_dependence} shows the $\beta$ uncertainty as a function of $\mathcal{M}$ for the two binning schemes.
The uncertainty decreases rapidly as $\mathcal{M}$ is increased up to approximately eight, while the changes at larger $\mathcal{M}$ are small and not monotonic, indicating that the sensitivity has largely saturated.
The optimized binning consistently yields a smaller uncertainty than the baseline.
The nominal choice $\mathcal{M}=8$ therefore already achieves a precision close to the best obtained with the larger numbers of bins considered in this study.
In practice, however, larger $\mathcal{M}$ implies fewer events per bin, which limits the precision with which the binned hadronic parameters can be determined from data.
The choice $\mathcal{M}=8$ also provides a balance between sensitivity and the statistical precision available in each bin.

\subsubsection{Effect of decay time acceptance}
\label{sec:decay_time_acceptance}

\begin{figure}[htb]
    \centering
    \includegraphics[width=0.9\textwidth]{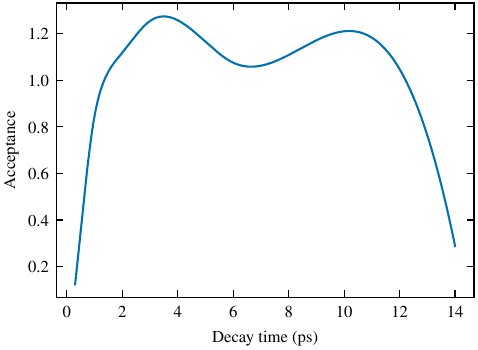}
    \caption{
        Representative relative decay time acceptance assumed in this study as a function of the $B$ decay time; its overall normalization is arbitrary.
    }
    \label{fig:dta}
\end{figure}

The relative decay time acceptance, defined up to an overall normalization, describes the reconstruction and selection response as a function of $B$ decay time and is included in the TD quality factor by weighting the time integration.
For this study, the representative decay time acceptance shown in Fig.~\ref{fig:dta} is assumed.
Its effect is found to be small.
This is expected because the same decay time weighting is applied to all Dalitz plot bins, so much of its effect cancels in the relative comparison of binning schemes.
A detailed comparison of including the decay time acceptance during the binning optimization and during the evaluation of the uncertainty improvement is presented together with the Dalitz plot efficiency study in Table~\ref{tab:effects}.

\subsubsection{Effect of Dalitz plot efficiency}

\begin{figure}[htb]
    \centering
    \includegraphics[width=0.47\textwidth]{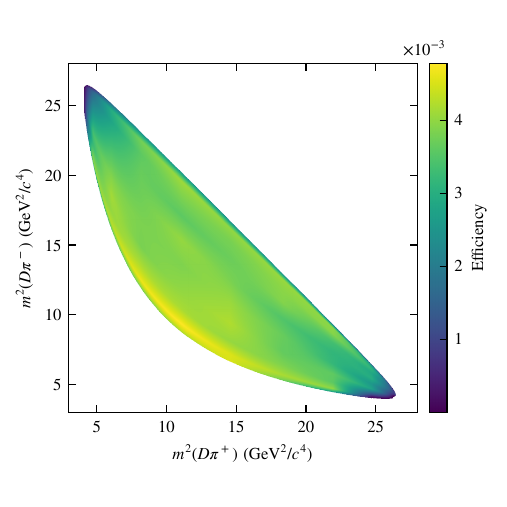}
    \hspace{0.03\textwidth}
    \includegraphics[width=0.47\textwidth]{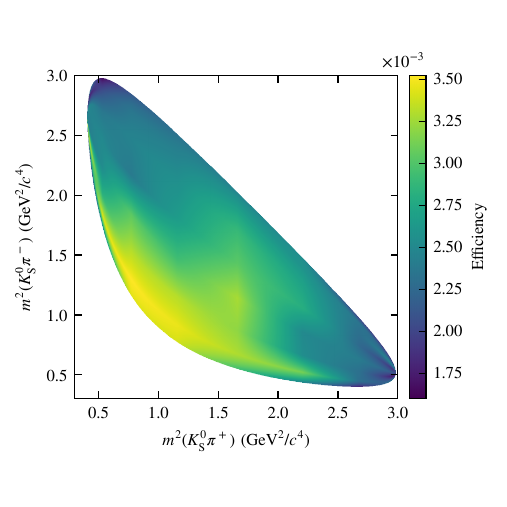}
    \caption{
        Representative efficiency maps assumed for the (left) $B$ and (right) $D$ Dalitz plots.
    }
    \label{fig:dlzeff}
\end{figure}

The Dalitz plot efficiency $\varepsilon(\Phi_B, \Phi_D)$ describes the probability for a signal decay at a given Dalitz plot position to be reconstructed and selected.
For this study, it is factorized as $\varepsilon(\Phi_B,\Phi_D)=\varepsilon_B(\Phi_B)\varepsilon_D(\Phi_D)$.
To incorporate the efficiency into the optimization and evaluation, the expected binned yields and interference terms are calculated from efficiency-weighted integrals.
For example, the efficiency-weighted $B$ decay parameters are
\begin{align}
    k_j         & = \int_{\mathcal{B}_j} \varepsilon_B(\Phi_B) \, |\mathcal{A}_{B \to \bar{D}}(\Phi_B)|^2 \, d\Phi_B \,, \\
    c_j + i s_j & = \frac{1}{\sqrt{k_j k_{-j}}}
    \int_{\mathcal{B}_j} \varepsilon_B(\Phi_B) \, \mathcal{A}_{B \to \bar{D}}^*(\Phi_B) \, \mathcal{A}_{\bar{B} \to D}(\Phi_B) \, d\Phi_B \,.
    \label{eq:eff_weighted}
\end{align}
The $D$ decay parameters $K_i$, $C_i$, and $S_i$ are calculated analogously, with the corresponding integrals weighted by $\varepsilon_D$.
For this study, representative efficiency maps with smooth, nonuniform variations across the $B$ and $D$ Dalitz plots are assumed, as shown in Fig.~\ref{fig:dlzeff}.
Such variations are typical of those reported in published Dalitz plot analyses; for example, Ref.~\cite{LHCb:2015} presents a phase-space-dependent efficiency function for $\Bz \to \Dzb\pip\pim$ decays.
The efficiencies can enter the study in two distinct ways: they can be included when optimizing the binning scheme, and they can be included when evaluating the uncertainty improvement of a fixed binning.
In the following, $\varepsilon_t$ denotes the decay time acceptance and $\varepsilon_{\rm DP}$ denotes the Dalitz plot efficiency.
Table~\ref{tab:effects} reports the TD quality-factor improvement for all combinations of these choices.

\begin{table}[htb]
    \centering
    \caption{
        Uncertainty improvement evaluated by $Q_{\rm TD}$ for TD-optimized binnings obtained and evaluated with different efficiency configurations.
        Values are given in percent.
    }
    \label{tab:effects}
    \small
    \setlength{\tabcolsep}{5pt}
    \begin{tabular}{cc|cccc}
        \hline
        \multicolumn{2}{c|}{Included in optimization} &
        \multicolumn{4}{c}{Included in evaluation}                                                                           \\
        $\varepsilon_t$                               & $\varepsilon_{\rm DP}$ &
        Neither                                       & $\varepsilon_t$        & $\varepsilon_{\rm DP}$ & Both               \\
        \hline
        No                                            & No                     & 9.7                    & 9.7  & 11.6 & 11.6 \\
        Yes                                           & No                     & 9.7                    & 9.7  & 11.6 & 11.6 \\
        No                                            & Yes                    & 9.0                    & 9.0  & 11.0 & 11.0 \\
        Yes                                           & Yes                    & 9.6                    & 9.6  & 11.5 & 11.5 \\
        \hline
    \end{tabular}
\end{table}

As anticipated in Sec.~\ref{sec:decay_time_acceptance}, including the decay time acceptance during the evaluation has a negligible effect: for any fixed optimized binning, the values in the ``Neither'' and $\varepsilon_t$ columns, or in the $\varepsilon_{\rm DP}$ and ``Both'' columns, are identical to the quoted precision.
Including the decay time acceptance during the optimization changes the improvement by at most $0.6$ percentage points among the configurations studied, which is small compared with the overall sensitivity gain.
In contrast, including the Dalitz plot efficiency during the evaluation changes the relative improvement of the nominal binning from $9.7\%$ to $11.6\%$.
This reflects the fact that the nonuniform efficiency affects the $\beta$ sensitivities of the equal-phase-difference and optimized binnings by different amounts, thereby increasing their relative difference.
For a given evaluation configuration, the spread among all four optimization configurations is at most $0.7$ percentage points, showing that the optimized sensitivity is stable against the representative efficiency variation.

\subsubsection{Effect of Dalitz plot resolution}
\label{sec:dalitz_resolution}

\begin{figure}[htb]
    \centering
    \includegraphics[width=0.65\textwidth]{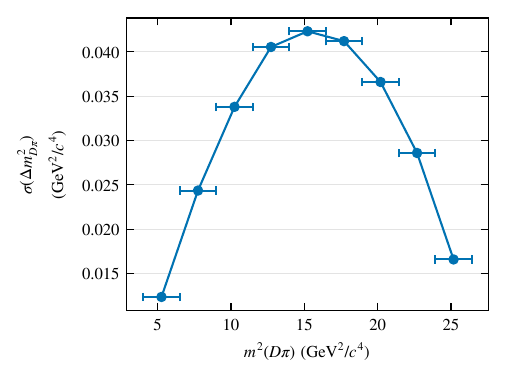}
    \caption{
        Representative position-dependent resolution profile assumed for each $m^2(D\pi)$ coordinate.
        The markers give the resolution widths $\sigma(\Delta m^2_{D\pi})$ at the centers of the intervals indicated by the horizontal bars, and the connecting line shows the linear interpolation used in the response model.
    }
    \label{fig:dalitz_resolution_profile}
\end{figure}

The nominal optimization does not account for finite resolution in the reconstructed Dalitz plot coordinates.
Such resolution can cause candidates to migrate across bin boundaries and may therefore reduce the sensitivity gain of an optimized binning if that gain relies on fine-scale variations across the Dalitz plot.
For this robustness test, a representative position-dependent resolution profile in each $m^2(D\pi)$ coordinate is assumed.
The profile is not intended to reproduce a particular detector response; it is used to test whether finite smearing changes the relative sensitivity gain.
The assumed profile is shown in Fig.~\ref{fig:dalitz_resolution_profile} and is implemented as a position-dependent, zero-centered response matrix.
This response is applied independently along the two $B$ Dalitz plot axes to the event and interference densities entering $k_j$, $c_j$, and $s_j$.

The effect on the nominal $\mathcal{M}=8$ TD-optimized binning is tested with toy pseudoexperiments containing approximately $13\,000$ signal events and no other experimental effects.
For both the equal-phase-difference baseline and the optimized binning, toys are generated and fitted using either the unsmeared or resolution-smeared binned parameters.
The ensemble means of the fit-reported $\beta$ uncertainties are summarized in Table~\ref{tab:dalitz_resolution}.

\begin{table}[htb]
    \centering
    \caption{
        Mean fit-reported $\beta$ uncertainties and their relative improvement for the nominal TD-optimized binning in toy pseudoexperiments without and with the representative $B$ Dalitz plot resolution.
        The uncertainties on the means and on the relative improvements are statistical.
    }
    \label{tab:dalitz_resolution}
    \small
    \setlength{\tabcolsep}{5pt}
    \begin{tabular}{lccc}
        \hline
        $B$ Dalitz resolution & Baseline ($^\circ$) & TD-optimized ($^\circ$) & Improvement (\%) \\
        \hline
        Not included          & $4.886 \pm 0.027$   & $4.433 \pm 0.022$       & $9.3 \pm 0.7$    \\
        Included              & $4.994 \pm 0.028$   & $4.526 \pm 0.023$       & $9.4 \pm 0.7$    \\
        \hline
    \end{tabular}
\end{table}

The corresponding $Q_{\rm TD}$ predictions for the relative improvement are $9.68\%$ without and $9.66\%$ with the resolution response.
The toy improvements differ by only $0.1$ percentage points and are statistically compatible.
Although the finite resolution increases the absolute $\beta$ uncertainties by approximately $2\%$, it does not significantly change the relative sensitivity gain from the optimized binning.
This small effect is consistent with the published $\Bz \to \Dzb\pip\pim$ Dalitz plot analysis of Ref.~\cite{LHCb:2015}, in which the invariant-mass resolution was found to have a negligible effect and was therefore not modeled.

\subsubsection{Effect of background}

\begin{figure}[htb]
    \centering
    \includegraphics[width=0.47\textwidth]{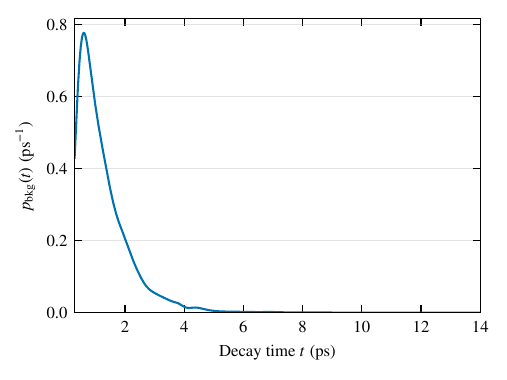}
    \hspace{0.03\textwidth}
    \includegraphics[width=0.47\textwidth]{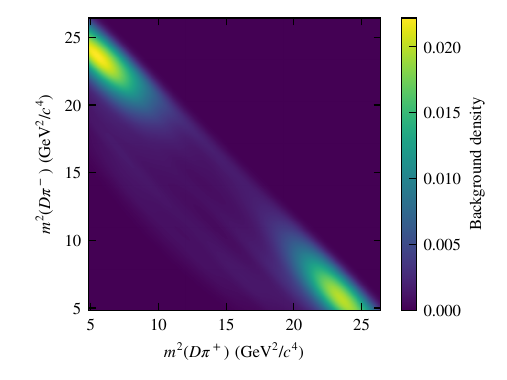}
    \caption{
        Representative background distributions assumed to construct $p_{\rm bkg}$: (left) decay time and (right) Dalitz plot variables.
    }
    \label{fig:bkg_kde}
\end{figure}

Backgrounds such as combinatorial background may affect the binning optimization, as the presence of background events dilutes the sensitivity to $\beta$ in each bin.
To study this effect, a background component is included in the PDF used to calculate the Fisher information.
The total PDF in each bin is expressed as
\begin{equation}
    p(t, i, j, q) = f_{\rm sig} \cdot p_{\rm sig}(t, i, j, q; \beta) + (1 - f_{\rm sig}) \cdot p_{\rm bkg}(t, i, j, q) \,,
\end{equation}
where $f_{\rm sig}$ is the signal fraction and $p_{\rm bkg}$ is the background PDF.
A representative background PDF $p_{\rm bkg}$ is assumed as the product of the decay time and Dalitz plot distributions shown in Fig.~\ref{fig:bkg_kde}.
The quality factor $Q^2$ is recalculated with the modified PDF, and the binning optimization is repeated.

\begin{figure}[htb]
    \centering
    \includegraphics[width=0.47\textwidth]{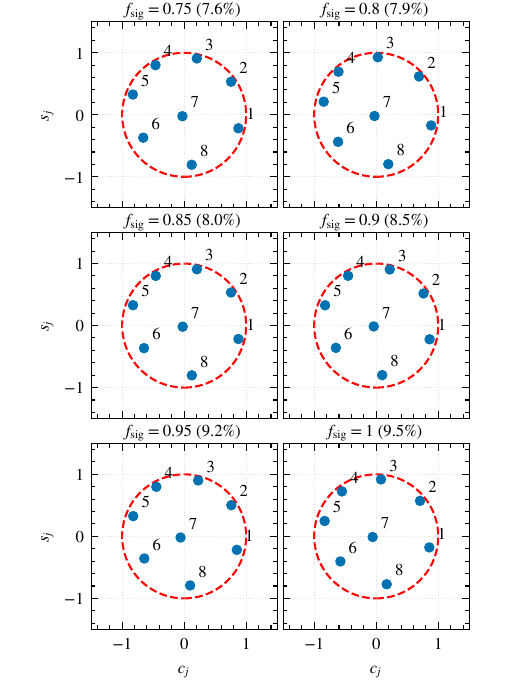}
    \hspace{0.03\textwidth}
    \includegraphics[width=0.47\textwidth]{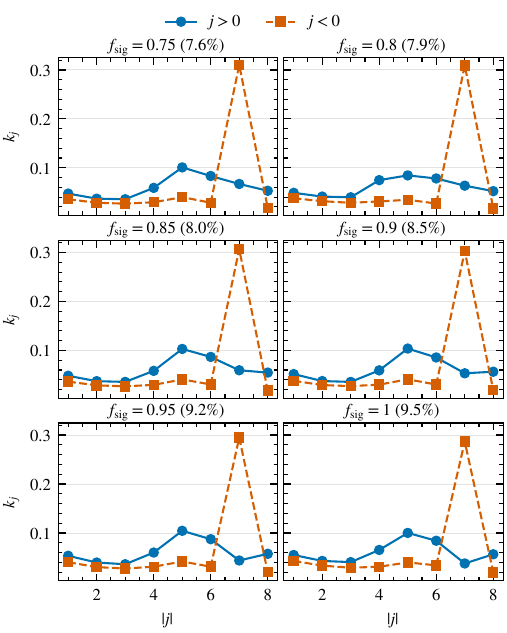}
    \caption{
        Distributions of (left) $c_j$--$s_j$ and (right) $k_j$ for the TD-optimized binning with different values of the signal fraction $f_{\rm sig}$.
        The percentages in the panel labels give the uncertainty improvement predicted by a subsequent background-free $Q_{\rm TD}$ evaluation using binned parameters calculated without experimental effects.
    }
    \label{fig:bkg_fsig}
\end{figure}

Figure~\ref{fig:bkg_fsig} shows the impact of varying the signal fraction $f_{\rm sig}$ on the optimized binning.
For the percentages displayed in the figure, each resulting binning is subsequently evaluated using the nominal background-free $Q_{\rm TD}$ with binned parameters calculated without experimental effects; thus, $f_{\rm sig}$ enters the optimization but not this evaluation.
The resulting $c_j$--$s_j$ and $k_j$ distributions change gradually as $f_{\rm sig}$ is varied, while the predicted uncertainty improvement increases overall with the signal fraction.
At $f_{\rm sig}=1$, the $9.5\%$ uncertainty improvement differs from the nominal $9.7\%$; this small variation is expected from stochastic optimization and does not affect the conclusions.
The benchmark value $f_{\rm sig} = 0.8$ is used.

\begin{figure}[htb]
    \centering
    \includegraphics[width=0.47\textwidth]{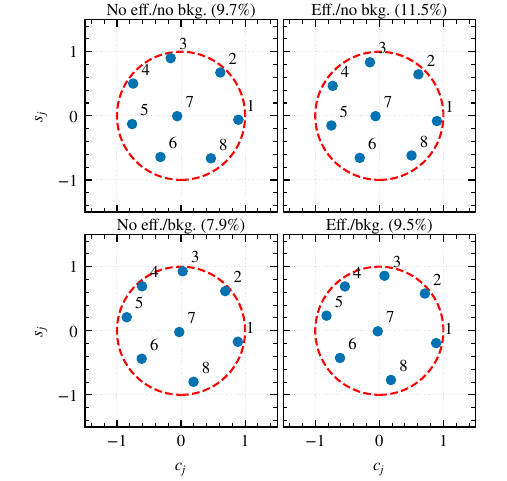}
    \hspace{0.03\textwidth}
    \includegraphics[width=0.47\textwidth]{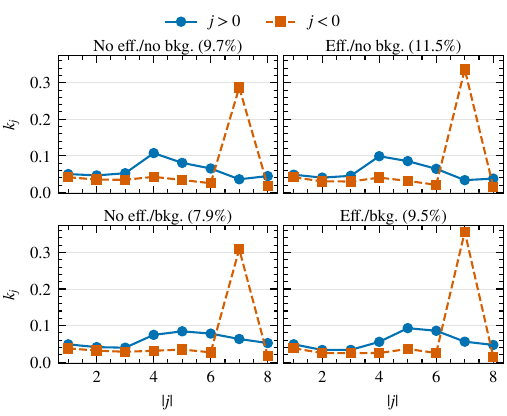}
    \caption{
        Comparison of (left) $c_j$--$s_j$ and (right) $k_j$ distributions for TD-optimized binnings obtained with the four combinations of the representative efficiencies (decay time acceptance and Dalitz plot efficiency together) and background component.
        The percentages in the panel labels give the uncertainty improvement predicted by a subsequent background-free $Q_{\rm TD}$ evaluation, with the efficiency configuration matched to that used in the optimization.
    }
    \label{fig:bkg_comparison}
\end{figure}

Figure~\ref{fig:bkg_comparison} compares four optimization configurations, formed by including or omitting the representative efficiencies (decay time acceptance and Dalitz plot efficiency together) and the background component in the TD method.
The resulting $c_j$--$s_j$ and $k_j$ parameters are broadly similar across the configurations.
The displayed percentages are obtained by subsequently evaluating each binning with the background-free $Q_{\rm TD}$, using the same efficiency configuration as in its optimization.
For a fixed efficiency choice, including the background at $f_{\rm sig}=0.8$ during the optimization changes the predicted uncertainty improvement by about two percentage points, without qualitatively altering the optimized parameters or the sensitivity gain.

\clearpage
\subsection{Verification with toy Monte Carlo}
\label{sec:toy_verification}

To verify that the $Q^2$-predicted improvements translate into actual gains in $\beta$ sensitivity, toy Monte Carlo studies are performed.
For each scenario, 250 pseudoexperiments are generated according to the time-dependent PDF of Eq.~(\ref{eq:pdf}) with the amplitude model used in the optimization.
A benchmark sample size of approximately $13\,000$ signal events is assumed in each pseudoexperiment.
The $\beta$ parameter is extracted from a maximum likelihood fit to each pseudoexperiment, in which the $B$ decay hadronic parameters $c_j$ and $s_j$ are fixed to their true values and the $D$ decay hadronic parameters $C_i$ and $S_i$ are fixed to their external inputs, while only $\beta$ is floated.
The generation uses the same values of $\beta$ and the mistag probability $\omega$ as assumed in the optimization described in Sec.~\ref{sec:annealing}.
The ensemble averages retain fits with valid minima and accurate covariance matrices, after rejecting fitted $\beta$ values more than eight fit-reported standard deviations from the generated value.

Table~\ref{tab:toy_verification} summarizes the results, showing the predicted improvement from $Q_{\rm TD}$ alongside the improvement observed in the toy fits.
The toy improvement is defined as $(\bar{\sigma}_{\rm base} - \bar{\sigma}_{\rm opt}) / \bar{\sigma}_{\rm base}$, where $\bar{\sigma}_{\rm base}$ and $\bar{\sigma}_{\rm opt}$ are the ensemble means of the fit-reported $\beta$ uncertainties for the baseline and optimized binnings, respectively.
The same nominal TD-optimized binning, obtained without efficiency effects, is used in both scenarios.
The scenarios differ only in the evaluation and toy generation: one is idealized, while the other includes the decay time acceptance and Dalitz plot efficiency described in Sec.~\ref{sec:effects}.

\begin{table}[htb]
    \centering
    \caption{
        Predicted and observed uncertainty improvements (\%) for the nominal TD-optimized binning with $\mathcal{M}=8$.
        The binning was optimized without efficiency effects in both rows; the rows differ only in the evaluation and toy generation.
        ``$Q$ prediction'' is evaluated from the TD quality factor, and ``toy result'' is constructed from the ensemble-mean fit-reported uncertainties.
    }
    \label{tab:toy_verification}
    \begin{tabular}{lcc}
        \hline
        Scenario                                            & $Q$ prediction (\%) & Toy result (\%) \\
        \hline
        No efficiency effects                               & 9.7                 & $9.3 \pm 0.7$   \\
        $\varepsilon_t$ and $\varepsilon_{\rm DP}$ included & 11.6                & $11.1 \pm 0.7$  \\
        \hline
    \end{tabular}
\end{table}

The good agreement between the $Q$-predicted and toy-fitted improvements confirms that the optimization identifies binning schemes that enhance the statistical sensitivity, and that the $Q^2$ quality factor is a reliable proxy for the $\beta$ uncertainty under the stated fit configuration.

\section{Summary}
\label{sec:summary}

The Dalitz plot binning optimization for the model-independent measurement of the CKM angle $\beta$ with $\Bz \to D\pip\pim$, $D \to \KS\pip\pim$ decays has been presented.
A Fisher information-based quality factor $Q^2$ is used to quantify the sensitivity retained after binning, and an optimized binning scheme for the $\Bz \to D\pip\pim$ Dalitz plot is found using simulated annealing.

Two optimization methods have been studied: a time-integrated (TI) method, in which the $B$ and $D$ contributions to $Q^2$ factorize, and a time-dependent (TD) method that fully accounts for $\Bz$--$\Bzb$ oscillations, decay time dependence, and flavour tagging.
The TI method, while simpler, predicts an overly optimistic improvement that does not fully reflect the structure of the $\beta$ measurement.
The TD method provides a more reliable estimate, yielding an uncertainty improvement of approximately $10\%$ relative to the equal-phase-difference baseline for the nominal $\mathcal{M}=8$ binning without experimental effects.

The predicted improvement has been validated through toy Monte Carlo studies, which show good agreement between the $Q^2$ prediction and the sensitivity gain in the stated fixed-hadronic-parameter $\beta$ fits.
The optimization is found to be robust against the experimental effects studied: including decay time acceptance in the evaluation has a negligible impact at the quoted precision; evaluating the fixed nominal scheme with the representative Dalitz plot efficiency changes the predicted relative improvement from $9.7\%$ to $11.6\%$; and the improvement obtained from toys with finite $B$ Dalitz plot resolution is statistically compatible with that obtained without resolution.
Including a representative background component during the optimization changes the uncertainty improvement subsequently predicted by the background-free TD quality factor by about two percentage points, without qualitatively altering the optimized parameters.

The final scheme recommended for future measurements at LHCb and Belle II is the TD-optimized symmetric binning with $\mathcal{M}=8$.
It is obtained in a signal-only configuration with uniform Dalitz plot efficiency, no Dalitz plot resolution smearing, and no background component; the robustness studies do not change the qualitative conclusion that this scheme improves the sensitivity.
The machine-readable bin definitions for the recommended scheme will be made publicly available alongside the final publication.
The optimization methodology is also applicable to other binned Dalitz plot analyses where the choice of binning affects the statistical sensitivity.

\section*{Acknowledgements}

The authors wish to thank their colleagues on the LHCb experiment for the fruitful and enjoyable collaboration that inspired this study.
In particular, they would like to thank Alex Bondar and Tim Gershon for helpful comments during this study.

\addcontentsline{toc}{section}{References}
\setboolean{inbibliography}{true}
\bibliographystyle{LHCb}
\small
\bibliography{references}

@article{Cabibbo:1963yz,
  author  = {Cabibbo, N.},
  title   = {{Unitary symmetry and leptonic decays}},
  journal = {Phys. Rev. Lett.},
  volume  = {10},
  pages   = {531},
  year    = {1963},
  doi     = {10.1103/PhysRevLett.10.531}
}

@article{Kobayashi:1973fv,
  author  = {Kobayashi, M. and Maskawa, T.},
  title   = {{CP violation in the renormalizable theory of weak interaction}},
  journal = {Prog. Theor. Phys.},
  volume  = {49},
  pages   = {652},
  year    = {1973},
  doi     = {10.1143/PTP.49.652}
}

@article{HFLAV,
  author        = {{Heavy Flavor Averaging Group}},
  title         = {{Averages of $b$-hadron, $c$-hadron, and $\tau$-lepton properties as of 2023}},
  journal       = {Phys. Rev. D},
  volume        = {113},
  number        = {1},
  pages         = {012008},
  year          = {2026},
  eprint        = {2411.18639},
  archiveprefix = {arXiv},
  doi           = {10.1103/x87q-tld5},
  note          = {with online updates at \url{https://hflav.web.cern.ch}}
}

@article{Latham:2008,
  author        = {Latham, T. and Gershon, T.},
  title         = {{A method to measure $\cos(2\beta)$ using time-dependent Dalitz plot analysis of $B^0 \to D_{CP}\pi^+\pi^-$}},
  journal       = {J. Phys. G},
  volume        = {36},
  pages         = {025006},
  year          = {2009},
  eprint        = {0809.0872},
  archiveprefix = {arXiv},
  doi           = {10.1088/0954-3899/36/2/025006}
}

@article{Gershon:2009qr,
  author        = {Gershon, T. and Poluektov, A.},
  title         = {{Double Dalitz plot analysis of the decay $B^0 \to D K^+\pi^-, D \to K_S\pi^+\pi^-$}},
  journal       = {Phys. Rev. D},
  volume        = {81},
  pages         = {014025},
  year          = {2010},
  eprint        = {0910.5437},
  archiveprefix = {arXiv},
  doi           = {10.1103/PhysRevD.81.014025}
}

@article{Giri:2003ty,
  author        = {Giri, A. and Grossman, Y. and Soffer, A. and Zupan, J.},
  title         = {{Determining $\gamma$ using $B^\pm \to D K^\pm$ with multibody $D$ decays}},
  journal       = {Phys. Rev. D},
  volume        = {68},
  pages         = {054018},
  year          = {2003},
  eprint        = {hep-ph/0303187},
  archiveprefix = {arXiv},
  doi           = {10.1103/PhysRevD.68.054018}
}

@article{Bondar:2005ki,
  author        = {Bondar, A. and Poluektov, A.},
  title         = {{Feasibility study of model-independent approach to $\phi_3$ measurement using Dalitz plot analysis}},
  journal       = {Eur. Phys. J. C},
  volume        = {47},
  pages         = {347},
  year          = {2006},
  eprint        = {hep-ph/0510246},
  archiveprefix = {arXiv},
  doi           = {10.1140/epjc/s2006-02590-x}
}

@article{Bondar:2008hh,
  author        = {Bondar, A. and Poluektov, A.},
  title         = {{The use of quantum-correlated $D^0$ decays for $\phi_3$ measurement}},
  journal       = {Eur. Phys. J. C},
  volume        = {55},
  pages         = {51},
  year          = {2008},
  eprint        = {0801.0840},
  archiveprefix = {arXiv},
  doi           = {10.1140/epjc/s10052-008-0600-z}
}

@article{Libby:2010nu,
  author        = {{CLEO Collaboration}},
  title         = {{Model-independent determination of the strong-phase difference between $D^0$ and $\bar{D}^0 \to K^0_{S,L} h^+ h^-$ ($h=\pi,K$) and its impact on the measurement of the CKM angle $\gamma/\phi_3$}},
  journal       = {Phys. Rev. D},
  volume        = {82},
  pages         = {112006},
  year          = {2010},
  eprint        = {1010.2817},
  archiveprefix = {arXiv},
  doi           = {10.1103/PhysRevD.82.112006}
}

@article{BESIII:2020:kspipi,
  author        = {{BESIII Collaboration}},
  title         = {{Determination of strong-phase parameters in $D \to K^0_{S,L} \pi^+ \pi^-$}},
  journal       = {Phys. Rev. Lett.},
  volume        = {124},
  pages         = {241802},
  year          = {2020},
  eprint        = {2002.12791},
  archiveprefix = {arXiv},
  doi           = {10.1103/PhysRevLett.124.241802}
}

@article{BESIII:2025:kspipi,
  author        = {{BESIII Collaboration}},
  title         = {{Updated model-independent measurement of the strong-phase differences between $D^0$ and $\bar{D}^0 \to K^0_{S, L} \pi^+ \pi^-$ decays}},
  journal       = {JHEP},
  volume        = {06},
  pages         = {086},
  year          = {2025},
  eprint        = {2503.22126},
  archiveprefix = {arXiv},
  doi           = {10.1007/JHEP06(2025)086}
}

@inproceedings{BESIII:2026:qc,
  author        = {Gilman, A.},
  title         = {{Quantum correlation of neutral charmed mesons at BESIII}},
  booktitle     = {Lepton Photon 2025},
  year          = {2026},
  eprint        = {2601.16768},
  archiveprefix = {arXiv}
}

@article{Rao:1945,
  author  = {Rao, C. R.},
  title   = {{Information and accuracy attainable in the estimation of statistical parameters}},
  journal = {Bull. Calcutta Math. Soc.},
  volume  = {37},
  pages   = {81--91},
  year    = {1945}
}

@article{Craik:2018,
  author        = {Craik, D. C. and Gershon, T. and Poluektov, A.},
  title         = {{Optimising sensitivity to $\gamma$ with $B^0 \to DK^+\pi^-$, $D \to K_{\rm S}^0\pi^+\pi^-$ double Dalitz plot analysis}},
  journal       = {Phys. Rev. D},
  volume        = {97},
  pages         = {056002},
  year          = {2018},
  eprint        = {1712.07853},
  archiveprefix = {arXiv},
  doi           = {10.1103/PhysRevD.97.056002}
}

@article{LHCb:2015,
  author        = {{LHCb Collaboration}},
  title         = {{Dalitz plot analysis of $B^0 \to \bar{D}^0 \pi^+\pi^-$ decays}},
  journal       = {Phys. Rev. D},
  volume        = {92},
  pages         = {032002},
  year          = {2015},
  eprint        = {1505.01710},
  archiveprefix = {arXiv},
  doi           = {10.1103/PhysRevD.92.032002}
}

@article{Bondar:2018:beta,
  author        = {Bondar, A. and Kuzmin, A. and Vorobyev, V.},
  title         = {{A method for model-independent measurement of the CKM angle $\beta$ via time-dependent analysis of $B^0 \to D\pi^+\pi^-$, $D \to K_S^0 \pi^+\pi^-$ decays}},
  journal       = {JHEP},
  volume        = {03},
  pages         = {195},
  year          = {2018},
  eprint        = {1802.00200},
  archiveprefix = {arXiv},
  doi           = {10.1007/JHEP03(2018)195}
}

@article{Grossman:1996:decay,
  author        = {Grossman, Y. and Worah, M. P.},
  title         = {{CP asymmetries in $B$ decays with new physics in decay amplitudes}},
  journal       = {Phys. Lett. B},
  volume        = {395},
  pages         = {241--249},
  year          = {1997},
  eprint        = {hep-ph/9612269},
  archiveprefix = {arXiv},
  doi           = {10.1016/S0370-2693(97)00068-3}
}

@article{Frings:2015eva,
  author        = {Frings, P. and Nierste, U. and Wiebusch, M.},
  title         = {{Penguin contributions to $CP$ phases in $B_{d,s}$ decays to charmonium}},
  journal       = {Phys. Rev. Lett.},
  volume        = {115},
  number        = {6},
  pages         = {061802},
  year          = {2015},
  eprint        = {1503.00859},
  archiveprefix = {arXiv},
  doi           = {10.1103/PhysRevLett.115.061802}
}

@article{Bondar:2005:beta,
  author        = {Bondar, A. and Gershon, T. and Krokovny, P.},
  title         = {{A method to measure $\phi_1$ using $\bar{B}^0 \to D h^0$ with multibody $D$ decay}},
  journal       = {Phys. Lett. B},
  volume        = {624},
  pages         = {1--10},
  year          = {2005},
  eprint        = {hep-ph/0503174},
  archiveprefix = {arXiv},
  doi           = {10.1016/j.physletb.2005.07.053}
}

@article{Belle:2016:beta,
  author        = {{Belle Collaboration}},
  title         = {{Measurement of the CKM angle $\varphi_1$ in $B^0 \to \bar{D}^{(*)0}h^0$, $\bar{D}^0 \to K_S^0\pi^+\pi^-$ decays with time-dependent binned Dalitz plot analysis}},
  journal       = {Phys. Rev. D},
  volume        = {94},
  pages         = {052004},
  year          = {2016},
  eprint        = {1607.05813},
  archiveprefix = {arXiv},
  doi           = {10.1103/PhysRevD.94.052004}
}

@article{Belle:2018:ambiguity,
  author        = {{BaBar and Belle Collaborations}},
  title         = {{First evidence for $\cos 2\beta > 0$ and resolution of the Cabibbo-Kobayashi-Maskawa quark-mixing Unitarity Triangle ambiguity}},
  journal       = {Phys. Rev. Lett.},
  volume        = {121},
  pages         = {261801},
  year          = {2018},
  eprint        = {1804.06152},
  archiveprefix = {arXiv},
  doi           = {10.1103/PhysRevLett.121.261801}
}

@article{Belle:2018,
  author        = {{BaBar and Belle Collaborations}},
  title         = {{Measurement of $\cos 2\beta$ in $B^0 \to D^{(*)} h^0$ with $D \to K_S^0 \pi^+ \pi^-$ decays by a combined time-dependent Dalitz plot analysis of BaBar and Belle data}},
  journal       = {Phys. Rev. D},
  volume        = {98},
  pages         = {112012},
  year          = {2018},
  eprint        = {1804.06153},
  archiveprefix = {arXiv},
  doi           = {10.1103/PhysRevD.98.112012}
}

@article{Kirkpatrick:1983,
  author  = {Kirkpatrick, S. and Gelatt, Jr., C. D. and Vecchi, M. P.},
  title   = {{Optimization by simulated annealing}},
  journal = {Science},
  volume  = {220},
  number  = {4598},
  pages   = {671},
  year    = {1983},
  doi     = {10.1126/science.220.4598.671}
}

\end{document}